

\magnification 1150
\baselineskip 14 pt
\newbox\Ancha
\def\gros#1{{\setbox\Ancha=\hbox{$#1$}
   \kern-.025em\copy\Ancha\kern-\wd\Ancha
   \kern.05em\copy\Ancha\kern-\wd\Ancha
   \kern-.025em\raise.0433em\box\Ancha}}
\font\bigggfnt=cmr10 scaled \magstep 3
 2
\font\bigfnt=cmr10 scaled \magstep 1
\font\ten=cmr8
\def\Par{\par\vskip 5 pt}
\def\eq#1{ \eqno({#1}) \qquad }
\vglue .3 in
\def\ie{{\it i.\ e.\ }}

\def \ni{\noindent}
\def\brad{|n_1\,j_1\,\epsilon_1\rangle}
\def\brai{ \langle n_2\,j_2\,\epsilon_2|}

\def\brakd{|n_1\,j_1\,\epsilon_1\rangle}
\def\braki{\langle n_2\,j_2\,\epsilon_2|}
\def\bi{\langle n_1\,j_1\,\epsilon_1|}
\def\bd{|n_2\,j_2\,\epsilon_2\rangle}

\ni{\bigggfnt  The Calculation of  Matrix Elements in\par\vskip5 pt \ni Relativistic Quantum Mechanics}\par
\vskip 10 pt

\ni {\bigfnt A.\ C.\ Ilarraza-Lomel\'{\i}, M. N. Vald\'es-Mart\'{\i}nez,  
\ni {\bigfnt A.\ L.\ Salas-Brito}\footnote{*}{\ten Corresponding author,
e-mail: asb@data.net.mx or asb@correo.azc.uam.mx}}\par
\noindent{\it Laboratorio de Sistemas Din\'amicos, Departamento de Ciencias B\'asicas, \par 
\noindent Universidad Aut\'onoma Metropolitana-Az\-ca\-pot\-zalco.\par
\noindent  Apar\-tado Pos\-tal 21--267, Co\-yoa\-c\'an, Mexico City, Distrito Federal C. P. 04000  M\'exico. }\Par

\noindent {\bigfnt R.\ P.\ Mart\'{\i}nez-y-Ro\-mero}\footnote{\dag}{\ten   e-mail: rodolfo@dirac.fciencias.unam.mx} 

\ni {\it Facultad de Ciencias, Universidad Nacional  Aut\'onoma  de M\'exico,\par\ni Apartado Postal 50-542, M\'exico City, Distrito Federal C.\ P.\ 04510 M\'exico.}\par
\vskip 8 pt
\noindent{\bigfnt H.\ N.\ N\'u\~nez-Y\'epez}\footnote{\ddag}{\ten  e-mail: nyhn@xanum.uam.mx}\par

\noindent{\it Departamento de F\'{\i}sica, Universidad Aut\'onoma Metropolitana-Iztapalapa}\par
\noindent{\it Apartado Postal 55--534, Iztapalapa, Distrito Federal C. P. 09340  M\'exico. }
\vskip 8 pt

\vskip 10 pt

\vskip 20pt
\vskip0.21truein

\centerline{\bigfnt Abstract.}\Par
\vskip0.035truein
\leftskip .3 in \rightskip .3 in \noindent    \Par
\vskip 2pt

Employing a relativistic version of a  hypervirial result,  recurrence relations for arbitrary  non-diagonal radial hydrogenic matrix elements have recently been obtained in Dirac relativistic quantum mechanics. In this contribution honoring Professor L\"owdin, we report on a new relation we have recently discovered between the matrix elements $\langle 2| r^\lambda |1 \rangle$ and  $\langle 2|\beta r^\lambda |1\rangle$---where $\beta$ is a  Dirac matrix and the numbers  distiguish between different radial  eigenstates--- that allow for a  simplification and hence for a more convenient way of expressing the recurrence relations.  We additionally derive another  relation that can be employed for simplifying  two center matrix element calculations in relativistic atomic or molecular calculations. 

\vfil

\vskip 10pt
\noindent PACS: 3.65.Ca\par

\eject
\noindent{\bf  Introduction}\Par	   

 The evaluation of expectation values  is always required for relating quantum calculations to experimental results in atomic or molecular physics. In most cases such expectation values  can be  expressed as matrix elements of powers of a radial coordinate $r$; this comes about since these powers  can be  regarded as either exact terms in a certain potential (as in the Lennard-Jones, the Casimir  or the London potentials) or as terms in a multipolar expansion of the interaction with the electromagnetic field [1--5]. Matrix elements of that sort can be also regarded  as  starting points of certain useful  approximation schemes, like variational or Hartree-Fock or configuration interaction methods, to which the late Professor L\"owdin  made important contributions [6--14]  and which admit relativistic extensions [15--17]. The treatment of electromagnetic  interactions in the realm of multiphoton transitions in very intense  laser fields usually needs a full quantum electrodynamics treatment [1], but this is  time consuming.   As a good approximation  we can use instead the relativistic Dirac quantum mechanical formalism [2]. One has only to remember the successes this theory has achieved in dealing with the hydrogen atom [18,19].  One can start  with the  known states of the relativistic hydrogen atom [18,20--23] and then proceed, as in non-relativistic quantum mechanics, to expand the states of interest  in terms of the former [24]. The problem is thus reduced to the evaluation of matrix elements of powers of $r$ between relativistic eigenstates of the hydrogen atom---a much simpler problem. The bad news are that these evaluations also become cumbersome. This calls for  techniques adroit for evaluating them. In nonrelativistic quantum mechanics these techniques in the form of algebraic methods, recurrence relations or clever uses of hypervirial theorems abound (see, for example,[25--31], but in the relativistic domain they are rather scarse (but see [4,32--34]).  Such lack of techniques is also manifest in the non existence of relations valid for two center matrix elements. 

The matrix elements of different powers of $r$ between Dirac eigenstates of the hydrogen atom referred to above  require the evaluation the following type of integrals

$$\eqalign{ 
\brai f(r) \brad  &= \int f(r) \left( F^*_2(r)F_1(r)+ G^*_2(r)G_1(r)\right) dr,    \cr
   \brai\beta f(r)\brad & = \int f(r) \left( F^*_2(r)F_1(r) - G^*_2(r)G_1(r) \right)   dr,          
} \eq{1}$$

\noindent where  $f(r)$ is any function of $r$, the kets $|n\,j\,\epsilon\rangle$ stand for
a bispinorial Dirac radial eigenstate of the hydrogen atom and the subscripts label different states. Any of such states, when projected on the $|r\rangle$
basis, become [2]

$$ \langle r|n\,j\,\epsilon\rangle =
 {1\over r}\left( \matrix{F_{n j \epsilon}(r)\cr \cr iG_{n j \epsilon}(r)}\right). 
\eq{2} $$

\noindent In equation (1) we used the shorthand $F_k=F_{n_kj_k\epsilon_k}(r)$ and $G_k=G_{n_kj_k\epsilon_k}(r)$, that  are called, respectively, the big and small components of the bispinor (2), and that are the solutions of the radial Dirac equation. This radial equation can be written [35]  as  $(H_k -E_k)\psi_k(r)=0$, where the $E_k$ are the energy eigenvalues, 

$$ H_k=c\alpha_r[p_r-i \beta\epsilon_k(j_k+1/2)/r] +\beta c^2+V_k(r), 
\eq{3}$$

\ni is the radial Dirac Hamiltonian, $k$ is just a label ---useful in what follows---, and $\beta= \hbox{ diag}(1, -1) $ is a Dirac matrix   (for a pedagogical discussion of the radial equation and of the hydrogen atom  in Dirac quantum mechanics see [36]), and [37--38] 

$$ \alpha_r= \pmatrix{0& -1\cr-1& 0}, \quad \hbox{ and }\quad p_r=-i\left({1\over r} +{d\over dr}\right). 
\eq{4}$$

\ni Writing the radial equation in a more explicit form, we have

$$ \left[\matrix{c^2+ V_k(r)-E_k& c{\epsilon_k\left(j_k+1/2\right)/ r} -c{d/ dr}\cr\cr
          c\epsilon_k\left(j_k+ 1/2\right)/r +c d/dr& -c^2+V_k(r)-E_k}\right] \left[\matrix{F_{n_kj_k\epsilon_k}(r)\cr\cr G_{n_kj_k\epsilon_k}(r)}\right]=0, 
\eq{5} $$

\noindent  where $n=0, 1, 2, \dots$ is the principal  quantum number, $j=1/2, 3/2, 5/2, \dots$ the total (orbital plus spin) angular momentum quantum number, $\epsilon\equiv(-1)^{j+l-1/2},\;$  $l\,(=j\pm 1/2$, according to whether $l$ refers to the big or to the small component of the hydrogenic spinor) is the orbital angular momentum quantum number, and $V_k(r)$ is any radial potential (of scalar type). Please note that here as in all of the paper we are using atomic units: $m_e=e=\hbar=1$. The quantum number $\epsilon$ equals $+1$ when $l=j+1/2$ and equals $-1$ when $l=j-1/2$ and it is related to the often used eigenvalue, $\kappa$, of the operator $ \beta(1+{\gros\Sigma}\cdot {\bf L})$  by $\kappa=-\epsilon(j+1/2)$, where  $\Sigma\equiv \hbox{diag}(\gros{\sigma}, \gros{\sigma})$ and $\gros{\sigma}=(\sigma_x,\sigma_y,\sigma_z)$ is the usual 3-vector spin operator. Notice also that, as in equations (2), (4), and (5) above, we can work in the somewhat easier to handle 2 dimensional subspace of the 4-dimensional Dirac operators.  This choice has no relevance whatsoever for the final results.  Closed forms for the integrals (1) or certain approximations thereof have been given in the Appendix of [38] and in   [34,39,40]  but even these become very cumbersome for many uses.  

On trying to overcome the complications mentioned above,   recurrence relations which can be used to compute  general ---not necessarily diagonal--- matrix elements between relativistic hydrogenic states of $r^\lambda$ have recently been obtained [37,38]. Given such recursions, it is only needed to evaluate at most 6 matrix elements, 3 for $r^\lambda$ and 3 more for $\beta r^\lambda$,  for obtaining  every other matrix element of $r^\lambda$ or of $\beta r^\lambda$ between hydrogenic states.

In this contribution honoring the memory of Professor L\"owdin and his work, we want to review these recurrence relations and to discuss one more relation, previously unnoticed to us,  which  can  greatly simplify the use of the  recurrence relations already reported. Other purpose of this contribution is to derive an hypervirial-like formula that can be useful for evaluating relativistic two center matrix elements.
\Par

\noindent{\bf  The previously known recurrence relations}\Par

The recurrence formulae have been obtained rederiving in the relativistic realm a  non-relativistic hypervirial result previously used to derive the Blanchard recursion relations between hydrogenic  matrix elements of   $r^\alpha$ terms [41,42]. The non-relativistic Blanchard relation has  been found so interesting that it has been generalized  [39]. We have to point out that our relativistic approach is totally different from Blanchard's; we start constructing an hypervirial and then proceed with several relativistic identities that have led us to the recurrence relations we are after [37,38]. This approach is inspired in a technique previously employed  for similar purposes in non-relativistic quantum mechanics [41].

To obtain a relativistically valid recurrence relation, let us first compute matrix elements of the radial function $\xi(r)=H_2 f(r)-f(r)H_1 $, where  $H_i$ is one of the Hamiltonians---these are really the same Hamiltonian just evaluated in any of the two states 1 or 2---appearing in Eq.\ (3),  to obtain, 

$$\eqalign{ 
E^-\langle n_2j_2 \epsilon_2|f(r)|n_1 &j_1\epsilon_1\rangle  = \langle n_2j_2\epsilon_2|H_2f(r)- f(r)H_1|n_1\,j_1\epsilon_1\rangle\cr& = -ic \langle n_2\,j_2 \epsilon_2|\alpha_r\left(f'(r)+ {\Delta^-_{21}\over 2r}\beta f (r)\right)|n_1\,j_1\epsilon_1\rangle,
} \eq{6}$$

\ni where  the primes are used to indicate $r$-derivatives and we  have introduced the symbols $ \Delta_{21}^\pm\equiv \epsilon_2(2j_2+1)\pm \epsilon_1(2j_1+1)$ and $E^\pm\equiv E_2\pm E_1$, ---in Eq.\ (4) only $\Delta_{21}^-$ and $E^-$ are used, $\Delta_{21}^+$ and $E^+$ will be used later on. Then computing  the matrix element  of $H_2 \xi(r)-\xi(r)H_1 $, we get [37,38]

$$\eqalign{ 
&(E^-)^2\brai  f(r)   \brad= \cr &\brai-{\Delta^{-}_{21} \over 2r^2} \beta f(r)-f''(r)-{\Delta^{-}_{21} \over 2r} f'(r) \beta- {\Delta^{-}_{21} \over r} f(r) \beta{d\over dr} + \cr &{\Delta_{21}^{+}\over 2r}  f'(r)\beta  +  \left({\Delta^{-}_{21}\over 2r}\right)^2 f(r) + 2 i \alpha_r\beta m\left(f'(r) + {\Delta^{-}_{21} \over 2r}\beta f(r)\right)\brad ,
} \eq{7}  $$

\ni  This formula is the relativistic extension of the  non relativistic second hypervirial introduced in a previous work [41]. 
Notice that though the non relativistic hypervirial  sufficed in determining the non relativistic recurrence relation it is not so in the relativistic case, we will also need the following two results: 

First, 

$$ \eqalign{
&E^+ E^-\braki r^\lambda\brakd =\braki c^2\left[{\Delta^-_{21}\Delta_{21}^+ \over 4}   + {\Delta^-_{21}\over 2} (1-\lambda) \beta\right] r^{\lambda -2} +\cr &
2Z\left[ic\alpha_r r^{\lambda -2}  (1-\lambda) - E^- r^{\lambda -1} \right] - E^+ \lambda ic\alpha_r r^{\lambda -1}\brakd.
} \eq{8}$$ 

\noindent were, for obtaining this relation we essentially repeat the steps leading to (7) excepting that, at the end, we evaluate the result on $H_2 \xi(r)\,+\,\xi(r)H_1 $. 

Second, we will also need 

$$\eqalign{
& \left( E^- -{4c^2\lambda\over \Delta^-_{21}}\right)\braki (-i\alpha_rr^{\lambda -1})\brakd=\cr
&\braki-c(\lambda -1 ) r^{\lambda -2}  - {4c\over \Delta^-_{21} }E^-r^\lambda  + c{\Delta^+_{21}\over 2}\beta r^{\lambda -2}\brakd ;
} \eq{9}$$

\noindent the detailed steps for obtaining (9) can be found in [38]. In Eqs.\ (8) and (9) and in most of what follows, we have explicitly used  $f(r)=r^{\lambda}$ and $V(r)=-Z/r$.  

The first of the recurrence relations we are after follows from eliminating the two terms, $2iE^+ \lambda  c \alpha_r r^{\lambda-1} $ and $2ic\alpha_r r^{\lambda-2}$, from (7) and (8). In this way we may get

$$ \eqalign{ 
c_0 \braki r^\lambda \brakd =\sum_{i=1}^{3} c_i\,\braki r^{\lambda -i} \brakd + \sum_{i=2}^{3} d_i\, \braki \beta r^{\lambda -i}\brakd, 
} \eq{10}$$

\ni where the numbers $c_i$,  $i=0,\dots 3,$ are given by

$$\eqalign{
c_0 & = {E^+(E^-)^2 \Delta^-_{21}\over D}, \cr
c_1 & = -{2 Z (E^-)^2 \Delta_{21}^-\over D + 4c^2},\cr
c_2 & = c^2{\Delta_{21}^-\Delta_{21}^+\over 4} -{c^2 \Delta_{21}^- 
\lambda (\lambda-1) E^+ \over D}, \cr
c_3& = {-2 Zc^2(\lambda -1)(\lambda -2)\Delta_{21}^-  \over D+ 4c^2 },
} \eq{11}$$

\ni and the numbers $d_i$, $i=2$ and 3, by

$$ \eqalign{
d_2 & = c^2{\Delta_{21}^-\over 2} \left[ (1-\lambda) + {\lambda E^+\Delta_{21}^+\over D}\right],\cr
d_3 & = {Z c^2(\lambda -1) \Delta_{21}^- \over Q}.
} \eq{12} $$

\ni where, for the sake of conciseness, we introduced the symbols

$$ D= \Delta_{21}^- E^- - 4 c^2\lambda, \qquad Q={  \Delta_{21}^- E^- - 4 c^2(\lambda-1)\over  \Delta_{21}^+}. 
\eq{13} $$

The second recurrence relation can be obtained from (8) through the simple but somewhat contrived  process explained in  [38], that yields

$$ \eqalign{
e_0  \braki \beta &r^\lambda \brakd = b_0 \braki r^{\lambda}\brakd + b_2 \braki r^{\lambda-2}\brakd \cr &+ e_1 \braki \beta r^{\lambda-1}\brakd  + e_2 \braki \beta r^{\lambda-2}\brakd,
} \eq{14} $$

\noindent where the numbers $b_r$ and $e_r$, $r=0, 2, 3,$ are given by

$$ \eqalign{
b_0=& 4\lambda\left[(E^-)^2 -4 c^4 \right], \cr
b_2=&c^2(1-\lambda)\left[(\Delta_{21}^{-})^2-4\lambda^2\right], \cr
             e_0=&2E^+ D,\cr
             e_1=&-4ZD,\cr
             e_2=& c^2{\Delta_{21}^+\over 2}[(\Delta_{21}^{-})^2-4\lambda^2].
} \eq{15} $$

\noindent Of course, the energy eigenvalues are those  of the relativistic hydrogen atom, namely

$$ E_a \equiv E_{n_a\, j_a}=  c^2\left(1+ {Z^2\alpha^2_F \over \left( n_a-j_a-1/2 +\sqrt{(j_a+1/2)^2-Z^2\alpha^2_F }\right)^2}\right)^{-1/2}, 
\eq{16} $$

\noindent  where $\alpha_F=1/c\simeq 1/137$ is the fine structure constant. Please note that contrary to what it was implied in [37,38] the recurrence relations (8) and (12) remain  valid even in the limit when $\Delta^{-}$ vanishes. \Par

\noindent{\bf  The new recurrence relations}\Par

Relations (9) and (12) can be useful for different computations, but they could be even more so if we could disentangle the matrix elements of  $r^a$ from the matrix elements of  $\beta r^b$. This can be achieved with the help of the following relationship 

$$ \eqalign{ 
\braki &{\Delta_{21}^+ \over 2r} f(r)\brakd= \cr&\braki \left[{\left(\Delta_{21}^+ + \Delta_{21}^-\right) \over 4rc^2}\left[E^+-2V(r)\right]-{1\over r}  \right] \beta f(r) + \beta f'(r)\brakd; 
} \eq{17}$$

\ni as follows from equation (8) by using the above illustrated technique but with different substitutions. Notice that we have reverted to an arbitrary radial function $f(r)$ and a generic radial potential $V(r)$. Now using, as we did before, the specific function  $f(r)=r^\lambda$ and the Coulomb potential $V(r)=-Z/r$, we obtain the {\sl useful  new relationship}

$$ g_0 \braki r^{\lambda}\brakd = j_0\braki \beta r^{\lambda}\brakd + j_1\braki \beta r^{\lambda-1}\brakd; 
\eq{18}  $$

\ni where 

$$ \eqalign{ 
g_0=& {\Delta^+_{21}/ 2}, \cr 
      j_0=& {E^+\over 4 c^2}(\Delta_{21}^+ + \Delta_{21}^-) + (\lambda-1), \cr 
      j_1=& {Z  \over 2 c^2}(\Delta_{21}^+ + \Delta_{21}^-). 
} \eq{19}$$ 

\ni Equation (18) is fundamental for the disentanglement of the recurrence relations (10) and (14).

Let us explain now the process for disentangling the matrix elements of $r^\lambda$ from those of $\beta r^\lambda$. We first obtain $\braki r^\lambda \brakd$ from equation (16) and from equation (14); by  equating the resulting expressions, we get the recurrence relation for terms of the form $\braki \beta r^\lambda\brakd$, as

$$ 
\eta_0 \braki \beta r^\lambda \brakd =\; \sum_{i=1}^3 \eta_i \braki\, \beta r^{\lambda-i} \,\brakd, 
 \eq{20}$$

\ni where the constants $\eta_a, \; a=0, 1,2, 3,$ appearing in (19) are

$$ \eqalign{
\eta_0=&  {E^+ D\over 2\lambda ((E^-)^2-4c^4)}-{R\over 2c^2}-{2\lambda \over \Delta_{21}^+},  \cr 
\eta_1=&  Z{R \over c^2 }- Z{D\over \lambda ((E^-)^2-4c^4)}, \cr 
\eta_2=& \left[{(\lambda-1)\over 2 \lambda} {4 \lambda^2 - (\Delta_{21}^-)^2 \over (E^-)^2-4c^4)} \right] 
\left[{E^+ R \over 4} + {c^2(\lambda - 2)\over \Delta_{21}^+ }  - {c^2 \Delta_{21}^- \over 4(\lambda-1)} \right], \cr   
\eta_3=& {Z (\lambda -1)R \over 4 \lambda }
\left[ {4 \lambda^2 - (\Delta_{21}^-)^2 \over (E^-)^2-4c^4} \right]. 
} \eq{21} $$

\ni Equation (20) is, together with the definitions (21), the new recursion for matrix elements involving just $\beta\, r^b$.

To obtain the recursion for the matrix elements of $r^a$, we start with equations (10) and (18). We first obtain the three $\braki \beta r^{\lambda-i}\brakd\; (i=0,1,2)$ terms from them;  next we succesively substitute them  into equation (14) and,  by juggling with the resulting equations, we are able to obtain

$$   \nu_0 \braki r^\lambda \brakd =\;\sum_{i=1}^5 \, \nu_i \, \braki r^{\lambda-i} \brakd 
 \eq{22}$$

\ni where the constants appearing in (22) are

$$ \eqalign{
\nu_0=& 2 (E^+E^-)^2 {Q \over Z c^2 (\lambda-1) }, \cr
\nu_1=& -8 E^+ (E^-)^2 {(Q + 6 c^2) \over c^2 (\lambda-1)}, \cr 
\nu_2=& {2 \lambda T \over Z } \left[ {\lambda E^+ \over (\lambda - 1) W}  
- Q - E^+ - {4c^2(\lambda-2) \over \Delta_{21}^+ + \Delta_{21}^-}\right] 
- {E^+ \over Z} D  \left[{U \over 2} -{2 c^2 \over R} \right] -  \cr 
& {J \over Z c^2 (\lambda - 1)} \left[ 8 Z^2 (E^-)^2 W + {c^2 \over 2} E^+ (E^-)^2 
{4 \lambda^2 - (\Delta_{21}^-)^2 \over Q W}  \right], \cr 
\nu_3= &- D \left[U - {4 c^2 \over R} \right] - 4 E^+ (\lambda - 2) Q W - {(E^-)^2 \left(4 \lambda^2 - (\Delta_{21}^-)^2\right) \over (\lambda - 1)}, \cr 
\nu_4=& {c^2 (\lambda - 1) \left(4 \lambda^2 - (\Delta_{21}^-)^2 \right) \over 2 Z } \left[ {\lambda E^+ \over (\lambda - 1) W} - Q - E^+ - {4c^2(\lambda-2) \over \Delta_{21}^+ + \Delta_{21}^-}\right] 
- \cr 
& 8Z (\lambda - 2)QW - {c^2 \Delta_{21}^+ \over 2 Z} \left(4 \lambda^2 - (\Delta_{21}^-)^2 \right) \left[ {U \over 4} - { c^2 \over R} \right], \cr 
\nu_5=& c^2 (\lambda -2) \left(4 \lambda^2 - (\Delta_{21}^-)^2 \right). 
} \eq {23} $$

\ni Again for the sake of conciseness, we have introduced the following definitions

$$ \eqalign{ 
Q =& {\Delta_{21}^- E^- -4c^2 (\lambda-1) \over \Delta_{21}^+}, \hskip 2.3cm    
    R = {\Delta_{21}^+ + \Delta_{21}^- \over \Delta_{21}^+}, \cr 
T =& \left( (E^-)^2 - 4c^4 \right), \hskip 3.2cm 
    W={\Delta_{21}^- E^- - 4 c^2 \lambda \over \Delta_{21}^- E^- -4 c^2 (\lambda-1)}, \cr
U =& \left[{\{\Delta_{21}^- E^- -4c^2(\lambda-1)\} [4 \lambda (\lambda -1) E^+ - \Delta_{21}^+ (\Delta_{21}^- E^- -4c^2\lambda)] \over (\Delta_{21}^- E^- -4c^2\lambda)(\lambda - 1) \Delta_{21}^+} \right].} \eq {24}$$ 

\ni Equation (22) is the recursion involving only matrix elements of $r^a$.

The new recurrence relations are equations (20) and (22), we also think that equation (19) can be found useful in certain applications. We must also pinpoint that the relationships derived are valid in general for complex values of the  exponent $\lambda$, as long as  $ \omega_1 + \omega_2 + |\lambda| \, > \, -1$, were the  numbers $\omega_i$ are defined as  $\sqrt{(j_i+1/2)^2-Z^2\alpha_F^2}$. This property was established in [38].\Par

\noindent{\bf  A  relation for two-center integrals.}\Par

The recurrence relations established above for relativistic hydrogenic states is sufficiently general for performing many calculations. It is not however useful for calculations involving matrix elements of a radial function between states represented by radial eigenstates corresponding to different potential functions, that is, the so-called two center matrix elements. It is the purpose of this section to present one such  relation for relativistic two-center matrix elements that can be regarded as a first step in the appropriate direction. This will be done using ideas developed in [38,41,43].\Par

General hypervirial results,  the virial theorem,  and other algebraic techniques [25,26,44] have been always very useful for calculating matrix elements, nonetheless they have been little used in relativistic calculations (but see [37,44,45]). We want to show here that they can have comparable importance than in non-relativistic quantum calculations. So, let us consider two  radial Dirac Hamiltonians [as in equation (3)] with two possibly different radial scalar potentials $V_1(r)$ and $V_2(r)$---we are thinking on potentials like those describing vibrational states in a molecule---

$$ \eqalign{H_1=&c\alpha_r[p_r-i \beta\epsilon_1 (j_1+1/2)/r] +\beta c^2+V_1(r),\cr
H_2=&c\alpha_r[p_r-i \beta\epsilon_2 (j_2+1/2)/r] +\beta c^2+V_2(r),
} \eq{25}$$

\ni a further assumption we make is that the two potentials are displaced from their respective equilibrium points by a constant quantity, \ie\  $r_1+r_2=a$, where $r_i$ is the equilibrium position of potential $V_i(r)$ and $a$ the displacement. Notice that, at difference with equation (3), here the numerical label in the Hamiltonians is not just convenient but has a definite physical meaning, for $H_2$ and $H_1$ refer to two in principle different systems.

 Taking the difference between the Hamiltonians (25), $H_1-H_2$, we get

$$ H_1= H_2 + ic\alpha_r\beta {\Delta_{21}^-\over 2r} - \left(V_2(r)-V_1(r) \right). \eq{26}$$ 

\ni On employing (25), we can inmediately evaluate the commutator

$$ [H_1, f_2(r)]=-ic\alpha_r {df_2(r)\over dr} \eq{27}$$

\ni where $f_2(r)$ is an arbitrary radial function and $[H, f(r)]$ stands for the commutator between $H$ and $f(r)$. We can calculate this commutator again, but now using expression (26) to get the alternative form

$$ [H_1, f_2(r)]= H_2f_2(r)-f_2(r) H_1 - \left(V_2-V_1 \right)f_2(r) + ic\alpha_r\beta 
{\Delta_{21}^-\over 2r}f_2(r). \eq{28}$$
 \Par

 If we now equal (28) with (27) and take matrix elements of the resulting expression between the two states, $\bi$ and $\bd$ ---which correspond to the different Hamiltonians (25)---  we directly obtain the hypervirial inspired relation

$$ \eqalign{(E_{2}&-E_{1})\bi f_2(r)\bd=\bi \left(V_2-V_1\right)f_2(r) \bd \cr -&ic\bi \alpha_r\left( f'_2(r)+\beta {\Delta_{21}^-\over 2r}f_2(r)\right)\bd; }
\eq{29} $$

\ni notice, however, that the energy eigenvalues in (29) refer to the different Hamiltonians  (26), and not to different states of the same Hamiltonian---as was the case analysed in the first section of the paper [compare with equation (6)]. In fact, if we consider the same potentials in equation (29), \ie\ we take $V_2(r)=V_1(r)$, we are also inmediately setting $E_1=E_2$  and thus we recover the  relationship (6), which is valid for just one center. 

The important point is  that equation (29) is an exact relation valid for the calculation of two center matrix elements, given as a function of the eigenenergies for any two scalar radial  potentials $V_i(r)$ in the Dirac equation, and thus it can be useful for deriving recurrence relations between such matrix elements. But this is still a work  in progress. \Par

\noindent{\bf  Acknowledgements.}\Par

 This work has been partially supported by CONACyT. This work owes a great deal to thoughtful comments and suggestions of our colleague and friend  C.\ Cisneros. ACI-L and MNV-M want to thank J. Morales for  what they have learnt from him.  F. C. Bonito, L. S. Micha, L. Bidsi, C. F. Quimo, C. Sabi, M. Botitas, F. C. Bieli, and all the gang  are also  acknowledged with thanks for their playful enthusiasm. This paper is dedicated to the memory of M. Osita.

\vfill
\eject

\noindent{\bf  References.}\Par

 \item {[1]} Cohen-Tannoudji, C.; Dupont-Roc, J.;  Grynberg, G. Atom-Photon Interactions; John Wiley: New York, 1992.

\item{[2]} Drake,  G. W. F. Atomic, Molecular, and Optical Physics Handbook; AIP: Woodbury New York,  1996. 

\item {[3]} Craig, D. P.;  Thirunamachandran, T. Molecular Quantum Electrodynamics; Academic: London, 1984.
 
\item {[4]} Wong, M. K. F.;   Yeh, H-Y.  Phys Rev A 1983, 27,  2300--2304. \par

\item {[5]} Dobrovolska, I. V.;  Tutik, R. S. Phys Lett A 1999,  260, 10--16.\par

\item {[6]} L\"owdin,  P. O. Phys Rev 1955, 97, 1474--1489.

\item {[7]} L\"owdin,  P. O. Phys Rev 1955, 97, 1490--1508.

\item {[8]} L\"owdin,  P. O.; Appel, K.; Phys Rev 1956, 103, 1746--1755.

\item {[9]} Shull, H.; L\"owdin,  P. O. Phys Rev 1958, 110, 1466--1467.

\item {[10]} L\"owdin,  P. O. Rev Mod Phys  1960, 32, 328--334.

\item {[11]} L\"owdin,  P. O. Rev Mod Phys  1962, 34, 80--86.

\item {[12]} L\"owdin,  P. O. Rev Mod Phys  1962, 34, 520--530.

\item {[13]} L\"owdin,  P. O.  Phys Rev  1965, 139, A357--A372.
 
\item {[14]} L\"owdin,  P. O. Rev Mod Phys  1967, 39, 259--287.

\item {[15]}  LaJohn, L. A.;  Christiansen, P. A.;  Ross, R. B.;  Atashroo, T.;   Ermler, W. C. J Chem Phys 1987, 87, 2812--2824. 

\item {[16]} Ley-Koo, E.; J\'auregui, R.; G\'ongora-T, A.; Bunge, C. F. Phys Rev A 1993, 47, 1761--1770. 

\item {[17]}  Chi, H.; Huang, K.;  Cheng, K. T. Phys Rev A 1991, 43, 2542--2545.

\item {[18]} Bethe, H. A.; Salpeter, E. E. Quantum Mechanics of One- and Two-Electron Atoms; Academic: New York, 1957. 

\item {[19]} Kim,  Y-K. Phys Rev 1965, 140, A1498--A1504.

\item {[20]} Mart\'{\i}nez-y-Romero, R. P. Am J Phys 2000,  68, 1050--1055.

\item {[21]} De Lange, O. L.;  Raab, R. E. Operator Methods in Quantum Mechanics; Clarendon: Oxford, 1991.

\item {[22]} Moss, R. E. Advanced Molecular Quantum Mechanics; Chapman and Hall: London, 1972.

\item {[23]} Davis, L. Phys Rev 1939, 56, 186--187. 

\item {[24]} Kim, Y-K. Phys Rev 1967, 154, 17--39.

\item {[25]} Fern\'andez, F. M.; Castro, E. A.  Hypervirial Theorems; Springer: New York,  1987.

\item {[26]} Fern\'andez, F. M.; Castro, E. A.  Algebraic Methods in Quantum Chemistry and Physics; CRC: Boca Raton, 1996.

\item {[27]} Picard, J.; de Irraza, C.;  Oumarou, B.;  Tran Minh, N.;  Klarsfeld, S. Phys Rev A 1991, 43, 2535--2537.

\item {[28]}  N\'u\~nez-Y\'epez,  H. N.;  L\'opez-Bonilla, J.;  Navarrete, D.;   Salas-Brito, A. L.  Int J Quantum Chem 1997,  62, 177--183.

\item {[29]} Morales, J; Pe\~na, J. J.; Portillo, P.; Ovando, G.; Gaftoi, V.  Int J Quantum Chem 1997,  65, 205--211.  

\item {[30]} Palting, P. Int. J Quantum Chem  1998, 67, 343--357. 

\item {[31]} Palting, P.; Villa, M.; Chiu, N. Y. J Mol Struct-Teo Chem 1999, 493, 51--62.
 
\item {[32]} Wong, M. K. F.;   Yeh, H-Y.  Phys Rev A 1983, 27,  2305--2310.

\item {[33]}  Kobus, J.;  Karkwowski, J.; Jask\'olski, W.  J Phys A: Math Gen 1987, 20, 3347--3352.

\item {[34]}  Shabaev, V. M. J Phys B: At Mol Opt Phys 1991,  24, 4479--4488. 

\item {[35]}    Grant, I. P.; in  Drake, G. W. F. (Editor)  Atomic, Molecular and Optical Physics Handbook; American Institute of Physics: Woodbury, 1996; Chapter 32.

\item {[36]} Mart\'{\i}nez-y-Romero, R. P. Am. J. Phys. 2000, 68, 1050--1055.

\item {[37]} Mart\'{\i}nez-y-Romero, R. P.;  N\'u\~nez-Y\'epez, H. N.;  Salas-Brito, A. L.  J Phys B: At Mol Opt Phys 2000,  33, L1--L8.

\item {[38]} Mart\'{\i}nez-y-Romero, R. P.;  N\'u\~nez-Y\'epez, H. N.;  Salas-Brito, A. L.  J Phys B: At Mol Opt Phys 2001,  34, 1261--1276.

\item {[39]} Shertzer, J. Phys Rev A 1991, 44, 2832--2835. 

\item {[40]}  Bessis, N.; Bessis, G.; Roux, D.   Phys Rev A 1985, 32, 2044--2050.

\item {[41]}  N\'u\~nez-Y\'epez, H. N.; L\'opez-Bonilla, J.; Salas-Brito, A. L.  J Phys B: At Mol Opt Phys  1995, 28, L525--L529.

\item {[42]}   Blanchard, P.;  J Phys B: At Mol Phys 1974, 7,  993--1005.

\item {[43]} Morales, J. Phys Rev A  1987, 36, 4101--4103.

\item{[44]} de Lange, O. L.;  Raab, R. E.  Operator Methods in Quantum Mechanics;  Clarendon: Oxford, 1991.

\item{[45]} Lucha, W.; Sch\"oberl F. F.; Phys Rev Lett 1990, 23, 2733--2735. 
\bye

\noindent{\bf References}\Par

D. P. Craig and T. Thirunamachandran, Molecular Quantum Electrodynamics, Academic Press , London 1984.

Chi Hsin-Chang, Huang Keh-Ning, K. T. Cheng, Phys. rev. A 43, 2542 (1991).

J. Picard, C. de Irraza, B. Oumarou, N. Tran Minh, S. Klarsfeld, PRA 43 2535

 Y-K.\ Kim Phys. Rev. 154, 17 (1967).

 Y-K.\ Kim, PHys.\ Rev.\ 140, A1498 (1965). 

C.\ Cohen-Tannoudji, J.\ Dupont-Roc, G.\ Grynberg, Atom-Photon Interactions, John Wiley  New York 1992.

E. Ley-Koo, R. J\'auregui, A. G\'ongora-T, C.F. Bunge, Phys. rev. A 47 1761 (1993).

\noindent 1. H.\ A.\ Kramers,  {\it  Quantum mechanics}, (North Holland,  Amsterdam, 1957).\par

\noindent 2.  P.\ W.\ Atkins, {\it Molecular Quantum Mechanics}, (Oxford, Clarendon, 1970). \par

\ni 3. J.\ Kobus, J.\ Karkwowski, and W.\ Jask\'olski, {\it J. Phys. A: Math. Gen. } {\bf 20}, 3347, (1987). \par

\ni 4. B.\ Moreno,  A.\ L\'opez-Pi\~neiro and R.\ H.\ Tipping, {\it J.\ Phys.\ A: Math.\ Gen.}, {\bf 24}, 385, (1991) .\par

\ni 5. H.\ N.\ N\'u\~nez-Y\'epez, J.\ L\'opez-Bonilla, D.\ Navarrete, and A.\ L.\ Salas-Brito, {\it Int.\ J.\ Quantum Chem.} {\bf 62}, 177, (1997).\par 

\ni 6. P.\ Blanchard, {\it  J. Phys. B: At. Mol. Phys.}, {\bf 7},  993, (1974). \par

\ni 7.  F.\ M.\ Fern\'andez and E.\ A.\ Castro, {\it Hypervirial Theorems}, (Springer, New York,  1987).\par

\ni 8. F.\ M.\ Fern\'andez and E.\ A.\ Castro, {\it Algebraic Methods in Quantum Chemistry and Physics}, (CRC, Boca Raton, 1996).\par

\ni 9. O.\ L.\ de Lange and R.\ E.\ Raab, {\it Operator Methods in Quantum Mechanics},  (Clarendon, Oxford, 1991).\par 

\ni 10. J.\ Morales, J.\ J.\ Pe\~na, P.\ Portillo, G.\ Ovando, and  V.\ Gaftoi,  {\it Int.\ J.\ of Quantum Chem.}, {\bf 65}, 205, (1997).\par

\ni 11.  H.\ N.\ N\'u\~nez-Y\'epez, J.\  L\'opez-Bonilla, and A.\ L.\ Salas-Brito, {\it  J.\ Phys.\ B: At.\ Mol.\ Opt.}, {\bf 28}, L525, (1995).\par

\ni 12. M.\ K.\ F.\ Wong  and H-Y.\ Yeh,  {\it Phys.\ Rev.\ A},  {\bf 27}, 2300, (1983).\par

\ni 13. I. V.\ Dobrovolska and R.\ S.\ Tutik, {\it Phys. Lett. A}, {\bf 260}, 10, (1999).\par

\ni 14. R.\ P.\ Mart\'{\i}nez-y-Romero, H.\ N.\ N\'u\~nez-Y\'epez, and A.\ L.\ Salas-Brito, {\it J.\ Phys.\ B: At.\ Mol.\ Opt.\ Phys.},  {\bf 33}, L367, (2000).\par

\ni 15. N.\ Bessis, G.\ Bessis, and D.\ Roux,  {\it Phys.\ Rev.\ A}, {\bf 32}, 2044, (1985).\par

\ni 16. V M Shabaev J.\ Phys.\ B: At.\ Mol.\ Opt.\ Phys.\ {\bf 24}, 4479 (1991).\par

\ni 17. M.\ K.\ F.\ Wong  and H-Y.\ Yeh,  {\it Phys.\ Rev.\ A},  {\bf 27}, 2305, (1983).\par

\ni 18. J.\ D.\ Bjorken and S.\ D.\ Drell,  {\it Relativistic Quantum Mechanics}, (Mac Graw-Hill, New York, 1964). \par

\ni 19.  I.\ P.\ Grant in G.\ W.\ F.\ Drake Editor, {\it Atomic, Molecular and Optical Physics Handbook}, (American Institute of Physics, Woodbury, 1996) Ch.\ 32.

\ni 20. R.\ E.\ Moss, {\it Advanced Molecular Quantum Mechanics}, (London, Chapman and Hall, 1972).\par

\ni 21. S.\ Pasternack, R.\ M.\ Sternheimer, {\it  J.\ Math.\ Phys.},  {\bf  3},  1280, (1962).\par

\ni 22.  Y. S. Kim, {\it Phys. Rev.} {\bf 154}, 17, (1967). \par

\ni 23. R.\ P.\ Mart\'{\i}nez-y-Romero, A.\ L.\ Salas-Brito and J.\ Salda\~na-Vega, {\it J. Math. Phys.},  {\bf 40}, 2324, (1999);

\ni 24. F.\ Constantinescu  and E.\ Magyari, {\it Problems in Quantum Mechanics}, (Oxford, Pergamon, 1971).\par

\ni 25.  R.\ P.\ Mart\'{\i}nez-y-Romero, A.\ L.\ Salas-Brito and J.\ Salda\~na-Vega, {\it J.\ Phys. A: Math. Gen.\ } {\bf 31}  L157 (1998).\par

\ni 26. U.\ Fano  and L.\ Fano,  {\it Physics of atoms and molecules}, (University of Chicago, Chicago, 1972).\par

\ni 27. L.\ Davies Jr.\ {\it Phys. Rev.}, {\bf  56}, 186, (1939).  \par

\ni  28. W.\ Magnus and  F.\ Oberhettinger, {\it Formulas and Theorems for the Special Functions of Mathematical Physics},  (Chelsea, New York, 1949).\par

\ni 29. R.\ Bluhm, V.\ A.\ Kostelecky, N.\ Rusell  Preprint hep-ph/0003223 (2000).\par

\vfill
\eject
\bye